\documentclass[showpacs,twocolumn,floatfix,superscriptaddress,prb,eqsecnum]{revtex4}
\usepackage{graphicx}
\usepackage{amsmath}
\begin{document}
\title{Quantum-to-classical crossover of mesoscopic conductance fluctuations}
\author{J. Tworzyd{\l}o}
\affiliation{Instituut-Lorentz, Universiteit Leiden, P.O. Box 9506, 2300 RA
Leiden, The Netherlands}
\affiliation{Institute of Theoretical Physics, Warsaw University, Ho\.{z}a 69, 00--681 Warsaw, Poland}
\author{A. Tajic}
\affiliation{Instituut-Lorentz, Universiteit Leiden, P.O. Box 9506, 2300 RA
Leiden, The Netherlands}
\author{C.W.J. Beenakker}
\affiliation{Instituut-Lorentz, Universiteit Leiden, P.O. Box 9506, 2300 RA
Leiden, The Netherlands}
\date{November 2003}
\begin{abstract}

We calculate the  system-size-over-wave-length ($M$) dependence
of sample-to-sample conductance fluctuations, 
using the open kicked rotator to model chaotic scattering
in a ballistic quantum dot coupled by two $N$-mode point
contacts to electron reservoirs. Both a fully quantum
mechanical and a semiclassical calculation are presented,
and found to be in good agreement.
The mean squared conductance fluctuations reach the universal
quantum limit of random-matrix-theory for small systems.
For large systems they increase $\propto M^2$ at
fixed mean dwell time $\tau_D \propto M/N$. The universal
quantum fluctuations dominate over the nonuniversal classical
fluctuations if $N < \sqrt{M}$. When expressed as a ratio of
time scales, the quantum-to-classical crossover is governed
by the ratio of Ehrenfest time and ergodic time.
\end{abstract}
\maketitle

\section{Introduction}

Sample-to-sample fluctuations of the conductance of
disordered systems have a universal regime, in which 
they are independent of the mean conductance. The 
requirement for these universal conductance fluctuations
\cite{Alt85,Lee85} is that the sample size should be
small compared to the localization length. The mean 
conductance is then much larger than the conductance 
quantum $e^2/h$.

The same condition applies to the universality of
conductance fluctuations in ballistic chaotic quantum 
dots \cite{Bar94,Jal94}, although there is no localization
in these systems. Random-matrix-theory (RMT) has the 
universal limit 
\begin{equation}
\lim_{N \rightarrow \infty} {\rm var}\, G=\frac{1}{8}
\end{equation}
for the variance of the conductance $G$ in units of $e^2/h$.
Here $N$ is the number of modes transmitted through each of
the two ballistic point contacts that connect the quantum
dot to electron reservoirs. Since the mean conductance
$\left< G \right>=N/2$, the condition for universality remains that
the mean conductance should be large compared to the
conductance quantum. 

In the present paper we will show that there is actually
an upper limit on $N$, beyond which RMT breaks down in a
quantum dot and the universality of the conductance fluctuations 
is lost. Since the width
$W$ of a point contact should be small compared to the 
size $L$ of the quantum dot, in order to have chaotic
scattering, a trivial requirement is $N \ll M$, where
$M$ is the number of transverse modes in a cross-section
of the quantum dot. (In two dimensions,
$N \simeq W/\lambda_{\rm F}$ and $M \simeq L/\lambda_{\rm F}$,
with $\lambda_{\rm F}$ the Fermi wavelength.) By considering
the quantum-to-classical crossover, we arrive at the more
stringent requirement 
\begin{equation}
1\ll N \ll \sqrt{M} e^{\lambda \tau_{\rm erg}/2},
\label{NllsqrtM}
\end{equation}
with $\lambda$ the Lyapunov exponent and $\tau_{\rm erg}$
the ergodic time of the classical chaotic dynamics. The 
requirement is more stringent than $N \ll M$ because, 
typically, $\lambda^{-1}$ and $\tau_{\rm erg}$ are both
equal to the time of flight $\tau_0$ across the system, so
the exponential factor in Eq.\ (\ref{NllsqrtM}) is not far from
unity.

Expressed in terms of time scales, the upper limit in Eq.\ 
(\ref{NllsqrtM}) says that $\tau_{\rm erg}$ should be larger
than the Ehrenfest time \cite{Vav02,Sil03}
\begin{equation}
\tau_E = \max \left[ 0,\lambda^{-1} \ln \frac{N^2}{M} \right].
\end{equation}
The condition $\tau_{\rm erg} > \tau_E$ which we find for the
universality of conductance fluctuations is much more 
stringent than the condition $\tau_{D} > \tau_E$ for the
validity of RMT found in other contexts. 
\cite{Lod98,Vav02,Sil03L,Jac02,Aga00,Sil03,Two03,Ale96,Ada03}
Here $\tau_D \approx (M/N) \tau_0$ is the mean dwell time in the quantum
dot, which is $\gg \tau_{\rm erg}$ in any chaotic system.

The outline of this paper is as follows. In Sec. II we
describe the quantum mechanical model that we use to calculate
${\rm var}\, G$ numerically, which is the same stroboscopic 
model used in previous investigations of the Ehrenfest time
\cite{Jac02,Two03,Gor03}. The data is interpreted semiclassically in 
Sec. III, leading to the crossover criterion (\ref{NllsqrtM}).
We conclude in Sec. IV.

\section{Stroboscopic model}

\begin{figure}
\includegraphics[width=7.5cm,height=7.5cm]{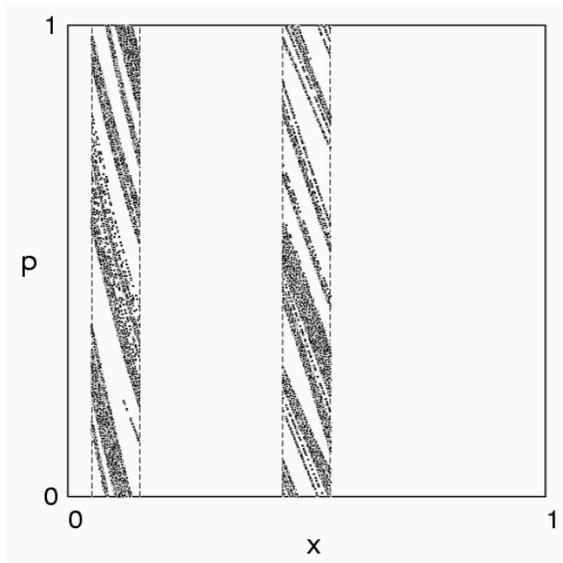}
\caption{  
Classical phase space of the open kicked rotator. The dashed lines
indicate the two leads (shown for the case $\tau_D=5$). Inside each lead we
plot the initial and
final coordinates of trajectories which are transmitted from the left
to the right lead after at most 3 iterations (with $K=7.5$). The points
cluster along narrow ``transmission bands''.
\label{fig_lead}
}
\end{figure}

The physical system we have in mind is a ballistic (clean)  quantum dot in a 
two-dimensional electron gas, connected by two ballistic leads to electron reservoirs.
 While the phase space of this system is four-dimensional, 
it can be reduced to two dimensions on a Poincar\'{e} surface of section.\cite{Bog92,Pra03}
The open kicked rotator 
\cite{Jac02,Two03,Gor03,Oss02,Bor91,Bor92,Fyo00} is a stroboscopic model
with a  two-dimensional phase space. We summarize how this model is constructed,
following Ref.\ \onlinecite{Two03}.

One starts from the closed system (without the leads). 
The kicked rotator describes a particle 
moving along a circle, kicked periodically at time intervals 
$\tau_{0}$. We set to unity the stroboscopic 
time $\tau_{0}$ and the Plank constant $\hbar$.
The stroboscopic time evolution of a wave function is given by the Floquet 
operator ${\cal F}$, which
can be represented by an $M\times M$ unitary symmetric matrix. 
The even integer $M$ defines the effective Planck constant $h_{\rm eff}=1/M$.
In the discrete coordinate 
representation ($x_m= m/M$, $m = 0,1, \ldots,M-1$) the matrix elements 
of ${\cal F}$ are given by
\begin{equation}
{\cal F}_{m' m} = M^{-1/2} e^{-i\pi/4} e^{i 2 \pi M S(x_{m'},x_m)}, \label{UPUdef}
\end{equation}
where $S$ is the map generating function,
\begin{equation}
S(x',x)= {\textstyle \frac{1}{2}} (x'-x)^{2} - (K/8\pi^2)(\cos 2\pi x' + \cos 2 \pi x ),
\end{equation}
and $K$ is the kicking strength.

The eigenvalues $\exp(-i\varepsilon_{m})$ of ${\cal F}$ define the 
quasi-energies  $\varepsilon_{m}\in (0,2\pi)$. The mean spacing $2\pi/M$ 
of the quasi-energies plays  the role of the mean level spacing $\delta$ 
in the quantum dot.

To model a pair of $N$-mode ballistic leads, we impose open boundary conditions 
in a subspace of Hilbert space represented by the indices $m_{n}^{(\alpha)}$. The
subscript $n=1,2,\ldots N$ labels the modes and  the superscript $\alpha=1,2$ labels
the leads. A $2N\times M$ projection matrix
$P$ describes the coupling to the ballistic leads. Its elements are
\begin{equation}
P_{nm}=\left\{\begin{array}{ll}
1&{\rm if}\;\;m=n\in\{m_{n}^{(\alpha)}\},\\
0&{\rm otherwise}.
\end{array}\right. \label{Wdef}
\end{equation}
The mean dwell time is $\tau_D = M/2N$ (in units of $\tau_0$).

The matrices $P$ and ${\cal F}$ together determine the quasi-energy dependent scattering matrix
\begin{equation}
S(\varepsilon)=P[e^{-i\varepsilon}-{\cal F}(1-P^{\rm T}P)]^{-1}{\cal F}P^{\rm T}.\label{Sdef}
\end{equation}
The symmetry of ${\cal F}$ ensures that $S$ is also symmetric, as it should be in the presence 
of time-reversal symmetry.
By grouping together the $N$ indices belonging to the same lead, the $2N\times 2N$ matrix $S$ can be decomposed into 4 sub-blocks containing the $N\times N$ transmission and reflection matrices,
\begin{equation}
S=\left(\begin{array}{cc}
r&t\\t'&r'
\end{array}\right).\label{Srt}
\end{equation}
The conductance $G$ (in units of $e^2/h$) follows from the Landauer formula
\begin{equation}
G={\rm Tr}\,tt^{\dagger}.\label{Gtt}
\end{equation}

The open quantum kicked rotator has a classical limit,
described 
by a map on the torus $\{x,p\;|\;{\rm modulo}\,1\}$. 
The classical phase space, including the leads, is shown in Fig.\ \ref{fig_lead}.
The map relates $x,p$ at  time $k$ to $x',p'$ at time $k+1$:
\begin{equation}
\label{mapdef}
p'=\frac{\partial }{\partial x'} S(x',x),\quad
p =-\frac{\partial}{\partial x}S(x',x).
\end{equation}
The classical mechanics becomes fully chaotic for $K\agt 7$, with Lyapunov
exponent $\lambda\approx \ln(K/2)$. For smaller $K$ the phase space is mixed,
containing both regions of chaotic and of regular motion. We will restrict 
ourselves to the fully chaotic regime in this paper.

\section{Numerical results \label{sec3}}

\begin{figure}
\includegraphics[width=8cm]{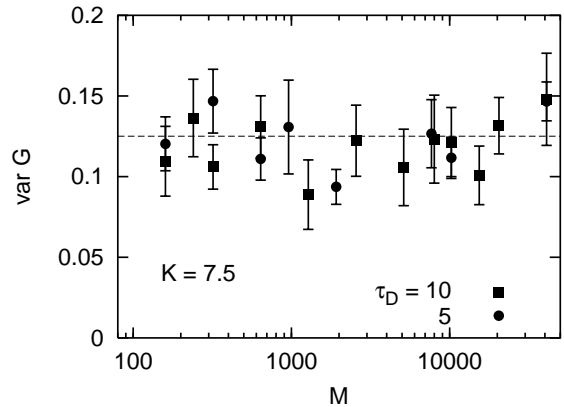}
\caption{  
\label{fig1}
Variance of the conductance fluctuations obtained numerically
by varying $\varepsilon$ with fixed lead positions. Error
bars indicate the scatter of values obtained for different 
lead positions. Results are shown as a function of
$1/h_{\rm eff}=M$, for two values of the dwell time
$\tau_D=M/2N$. The dashed line is the RMT prediction
${\rm var}\, G =\frac{1}{8}$.
}
\end{figure}

To calculate the conductance (\ref{Gtt})
we need to invert
the $M\times M$ matrix between square brackets in Eq. (\ref{Sdef}).
We do this numerically using an iterative procedure. \cite{Two03}
The iteration can be done efficiently using the fast-Fourier-transform
algorithm to calculate the application of $\cal F$ to a vector.
The time required to calculate
$S$ scales as $M^2\ln M$, which for large $M$ is quicker than the $M^3$
scaling of a direct inversion.
The memory requirements scale as $M$, because we need not store the
full scattering matrix to obtain the conductance.

We distinguish two types of mesoscopic fluctuations in the conductance. The first type
appears upon varying the quasi-energy $\varepsilon$ for a given scattering matrix 
$S(\varepsilon)$. Since these fluctuations have no classical analogue (the classical
map (\ref{mapdef}) being $\varepsilon$-independent), we refer to them as quantum 
fluctuations. The second type appears upon varying the position of the leads, so 
these involve variation of the scattering matrix at fixed $\varepsilon$. We refer
to them as sample-to-sample fluctuations.
They have both a quantum mechanical component and a classical analogue. One could 
introduce a third type of fluctuations, involving both variation of $\varepsilon$
and of the lead positions. We have found (as expected) that these are statistically 
equivalent to the  sample-to-sample fluctuations at fixed $\varepsilon$, so we need
not distinguish between fluctuations of type two and three. 

\begin{figure}[h]
\includegraphics[width=8cm]{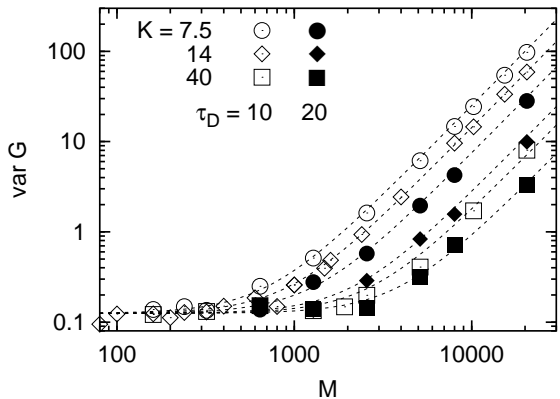}
\caption{  
Same as Fig.\ \ref{fig1}, but now for an ensemble in which
the lead positions and the quasi-energy are both varied.
The dashed lines are the sum of the RMT value (\ref{Vrmt}) and the 
classical result (\ref{Vcl}). Results are shown for three values of
the kicking strength $K$. Open symbols are for the dwell time $\tau_D=10$
and closed ones for $\tau_D=20$.
\label{fig2_a}
}
\end{figure}

We have calculated the variance ${\rm var}\, G=\left< G^2 \right>-\left< G \right>^2 $
of the conductance, either by varying $\varepsilon$
at fixed lead positions (quantum fluctuations) or by varying both $\varepsilon$ and
lead positions (sample-to-sample fluctuations).
Since the quantum interference
pattern is completely different only for energy variations
of order of the Thouless energy $1/ \tau_D$, we choose a number $\tau_D$
of equally spaced values of $\varepsilon$ in the interval $(0,2\pi)$.
We take $10$
different lead positions, randomly located at the $x$-axis in Fig.\ \ref{fig_lead}.
To investigate the quantum-to-classical crossover, we
change $h_{\rm eff}=1/M$ while keeping the dwell time
$\tau_D = M/2N$ constant. The results are plotted 
in Figs.\ \ref{fig1} and \ref{fig2_a}.

\section{Interpretation}

We interpret the numerical data by assuming that the variance of the
conductance is the sum of two contributions: a universal quantum
mechanical contribution ${\cal V}_{\rm RMT}$ given by random-matrix theory
and a nonuniversal quasiclassical contribution ${\cal V}_{\rm cl}$
determined by sample-to-sample fluctuations in the classical
transmission probabilities.

The RMT contribution equals \cite{Bar94,Jal94}
\begin{equation}
{\cal V}_{\rm RMT} = {\textstyle \frac{1}{8}},
\label{Vrmt}
\end{equation}
in the presence of time-reversal symmetry. The classical contribution is 
calculated from the classical map (\ref{mapdef}), by determining the probability
$P_{1\rightarrow 2}$ of a particle injected randomly through lead $1$ to
escape via lead $2$. Since the conductance is given semiclassically by
$G_{\rm cl} = N P_{1\rightarrow 2}$, we obtain
\begin{equation}
{\cal V}_{\rm cl}= N^2\, {\rm var}\, P_{1\rightarrow 2}.
\label{Vcl}
\end{equation}

We plot ${\rm var}\, G = {\cal V}_{\rm RMT}+{\cal V}_{\rm cl} $ in Fig.\ \ref{fig2_a} 
(dashed curves), for comparison with the
results of our full quantum mechanical calculation. The agreement
is excellent. 

\begin{figure}
\includegraphics[width=8cm]{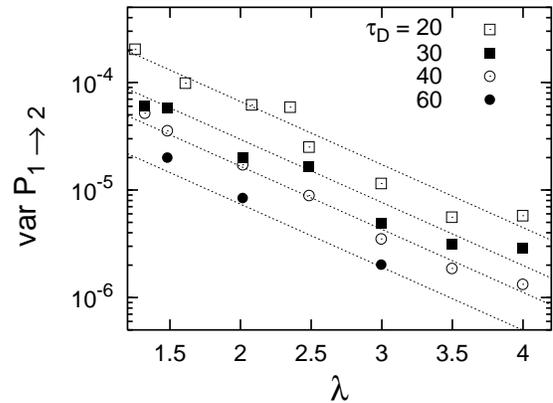}
\caption{ Variance of the classical fluctuations of the
transmission probability $P_{1 \rightarrow 2}$ upon changes of lead positions,
calculated numerically from the map (\ref{mapdef}).
The data is shown for four values of the dwell time $\tau_D$,
as a function of the Lyapunov exponent $\lambda = \ln(K/2)$.
The dotted lines are the analytical prediction (\ref{varPestimate}), with
fit parameters
$c = 1.6$ and $\tau_{\rm erg} = 0.68$ (the same for all data sets).
\label{fig3}
}
\end{figure}

We now would like to investigate what ratio of time scales governs
the crossover from quantum to classical fluctuations.

To estimate the magnitude of the sample-to-sample fluctuations in the classical
transmission probability, we use results from Ref.\ \onlinecite{Sil03}. There 
it was found that the starting points (and end points) of transmitted trajectories are not
homogeneously distributed in phase space. Instead, they cluster together in
nearly parallel, narrow bands. These transmission bands are clearly visible in
Fig.\ \ref{fig_lead}. The largest band has an area 
$A_{\rm max}=A_{0}e^{-\lambda\tau_{\rm erg}}$ determined by the ergodic time 
$\tau_{\rm erg}$. This is the time
required for a trajectory to explore the whole accessible phase space. The
values of $\tau_{\rm erg}$ and $A_{0}$ depend on the degree of collimation of
the beam of trajectories injected into the system. \cite{Sil03} For our model,
without collimation, one has $\tau_{\rm erg}$ of order unity (one stroboscopic
period) and $A_{0}\simeq(N/M)^{2}$. The typical transmission band has an area
$A_{0}e^{-\lambda\tau_{D}}$ which is exponentially smaller than $A_{\rm max}$
(since $\tau_{D}=M/2N\gg\tau_{\rm erg}$).

As the position of the lead is moved around, transmission bands move into and
out of the lead. The resulting fluctuations in the transmission probability
$P_{1\rightarrow 2}$ are dominated by the largest band. Since there is an
exponentially large number $e^{\lambda\tau_{D}}$ of typical bands, their
fluctuations average out. The total area in phase space of the lead is $A_{\rm
lead}=N/M$, so we estimate the mean squared fluctuations in $P_{1\rightarrow
2}$ at
\begin{equation}
{\rm var}\,P_{1\rightarrow 2}\simeq(A_{\rm max}/A_{\rm lead})^{2}
= c (N/M)^{2}e^{-2\lambda\tau_{\rm erg}},\label{varPestimate}
\end{equation}
with $c$ and $\tau_{\rm erg}$ of order unity.
We have tested this functional dependence numerically for the map 
(\ref{mapdef}), and find a reasonable agreement (see Fig.\ \ref{fig3}).
Both the exponential dependence on $\lambda$ and the quadratic dependence on
$\tau_D=M/2N$ are consistent with the data. We find $\tau_{\rm erg} = 0.68$
of order unity, as expected.

\begin{figure}[h]
\includegraphics[width=8cm]{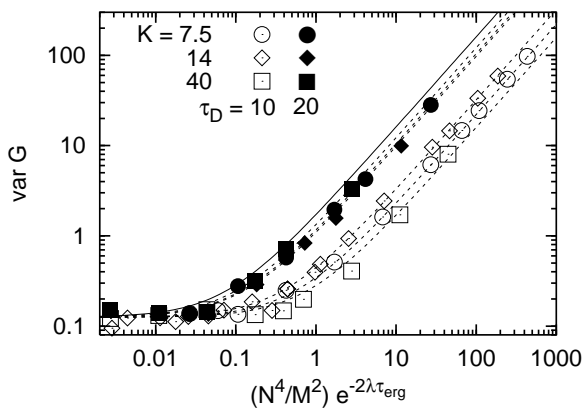}
\caption{  
Same data as in Fig.\ \ref{fig2_a} 
rescaled to show the approach to a single limiting curve
in the large dwell time limit.
The solid line is calculated from Eq.\ (\ref{varForm}),
with the same parameters $c=1.6$, $\tau_{\rm erg}=0.68$
as in Fig.\ \ref{fig3}.
\label{fig2_b}
}
\end{figure}

Eqs. (\ref{Vcl}) and (\ref{varPestimate}) imply 
\begin{equation}
{\rm var}\, G ={\textstyle \frac{1}{8}} + c (N^4/ M^2) e^{-2\lambda\tau_{\rm erg}}.
\label{varForm}
\end{equation}
In Fig.\ \ref{fig2_b}
we plot the same data as in Fig.\ \ref{fig2_a}, but now 
as a function of $(N^4/ M^2) e^{-2\lambda\tau_{\rm erg}}$.
We see that the functional dependence (\ref{varForm})
is approached for large dwell times.

The quantum fluctuations of RMT dominate over the classical fluctuations if
$N^{2}\,{\rm var}\,P_{1\rightarrow 2}\ll 1$. Using the estimate
(\ref{varPestimate}), this amounts to the condition
\begin{equation}
\tau_{\rm erg}>{\rm max}\left[ 0,\lambda^{-1}\ln(N^{2}/M) \right] \equiv
\tau_{E}\label{crossovercondition}
\end{equation}
that the ergodic time exceeds the Ehrenfest time. Notice that condition
(\ref{crossovercondition}) is always satisfied if $N^{2}<M\equiv 1/h_{\rm
eff}$. This agrees with the findings of Ref.\ \onlinecite{Sil03}, that the breakdown
of RMT starts when  $N \gtrsim \sqrt{M}$.

\section{Conclusions}

In summary, we have presented both a fully quantum mechanical and a semiclassical
calculation of the quantum-to-classical crossover from
universal to non-universal conductance fluctuations. The two 
calculations are in very good agreement, without any adjustable
parameter (compare data points with curves in Fig. \ref{fig2_a}).
We have also given an analytical approximation to the numerical 
data, which allows us to determine the parametric dependence
of the crossover.

We have found that universality of the conductance fluctuations
requires the ergodic time $\tau_{\rm erg}$ to be larger
than the Ehrenfest time $\tau_E$. This condition is much
more stringent than  the condition that the dwell time
$\tau_D$ should be larger than $\tau_E$, found previously
for universality of the shot noise in a quantum dot.
\cite{Aga00,Sil03,Two03}
The universality of the excitation gap in a quantum dot
connected to a superconductor is also governed by the ratio
$\tau_D/\tau_E$ rather than $\tau_{\rm erg}/\tau_E$ \cite{Lod98,Vav02,Sil03L,Jac02},
as is the universality of the weak localization effect. 
\cite{Ale96,Ada03}
These two properties have in common that they represent
ensemble averages, rather than sample-to-sample fluctuations.

We propose that what we have found here for the conductance
is generic for other transport properties: That the breakdown
of RMT with increasing $\tau_E$ occurs when $\tau_E > \tau_D$
for ensemble averages and when $\tau_E > \tau_{\rm erg}$ for
the fluctuations. This has immediate experimental consequences,
because it is much easier to violate the condition $\tau_E > \tau_{\rm erg}$
than the condition $\tau_E > \tau_D$.

To test this proposal, an obvious next step would be to determine
the ratio of time scales that govern the breakdown of universality
of the fluctuations in the superconducting excitation gap. The numerical
data in Refs.\ \onlinecite{Gor03} and \onlinecite{Kor03} was interpreted
in terms of the ratio $\tau_E/\tau_D$, but an alternative description
in terms of the ratio $\tau_E/\tau_{\rm erg}$ was not considered.

One final remark about the distinction between classical and quantum 
fluctuations, explained in Sec.\ \ref{sec3}.
It is possible to suppress the classical fluctuations entirely,
by varying only the quasi-energy at fixed lead positions. In that case we 
would expect the breakdown of universality to be governed by $\tau_D/\tau_E$
instead of $\tau_{\rm erg}/\tau_E$. Our numerical data (Fig. \ref{fig1})
does not show any systematic deviation from RMT, probably because we could
not reach sufficiently large systems in our simulation.

{\em Note added}: Our final remark above has been criticized by Jacquod and Sukhorukov
[\onlinecite{Jac03}]. They argue that the numerical data of Fig.\ \ref{fig1}
(and similar data of their own) does not show any systematic deviation
from RMT because quantum fluctuations remain universal if $\tau_E > \tau_D$.
Their argument relies on the assumption that the effective RMT of 
Ref.\ \onlinecite{Sil03} holds not only for the classical fluctuations (as we 
assume here), but also for the quantum fluctuations. The effective RMT says that
quantum fluctuations are due to a number $N_{\rm eff} \approx N e^{-\tau_E / \tau_D}$
of transmission channels with an RMT distribution. Universality of the quantum
fluctuations is then guaranteed even if  $N_{\rm eff} \ll N$, as long
as  $N_{\rm eff}$ is still large compared to unity.

This line of reasoning, if pursued further, contradicts the established theory
\cite{Ale96,Ada03} of the $\tau_E$ dependence of weak localization. RMT says
that the weak localization correction $\delta G=-\frac{1}{4}$ is independent
of the number of channels \cite{Bar94,Jal94}. Validity of the effective
RMT at the quantum level would therefore imply that weak localization 
remains universal if $\tau_E > \tau_D$, as long as $N e^{-\tau_E / \tau_D} \gg 1$.
This contradicts the result $\delta G = \frac{1}{4} e^{-\tau_E / \tau_D}$
of Refs. \onlinecite{Ale96} and \onlinecite{Ada03}. 
 
\acknowledgments
This work was supported by the Dutch Science Foundation NWO/FOM. J.T. 
acknowledges the financial support provided through the European 
Community's Human Potential  Programme under contract HPRN--CT--2000-00144, 
Nanoscale Dynamics.


\begin{thebibliography}{99}

\bibitem{Alt85} B. L. Altshuler, JETP Lett. {\bf 41}, 648 (1985).
\bibitem{Lee85} P. A. Lee and A. D. Stone, Phys.\ Rev.\ Lett.\ {\bf 55}, 1622 (1985).
\bibitem{Bar94} H. U. Baranger and P. A. Mello,  Phys.\ Rev.\ Lett.\ {\bf 73}, 142 (1994).
\bibitem{Jal94} R. A. Jalabert, J.-L. Pichard, and C. W. J. Beenakker, Europhys.\ Lett.\ {\bf 27}, 255 (1994).
\bibitem{Vav02} M. G. Vavilov and A. I. Larkin, Phys.\ Rev.\ B {\bf 67}, 115335 (2003).
\bibitem{Sil03} P. G. Silvestrov, M. C. Goorden, and C. W. J. Beenakker,
Phys.\ Rev.\ B {\bf 67}, 241301 (2003).
\bibitem{Lod98} A. Lodder and Yu. V. Nazarov, Phys. Rev. B {\bf{58}}, 5783 (1998).

\bibitem{Sil03L} P. G. Silvestrov, M. C. Goorden, and C. W. J. Beenakker,
Phys.\ Rev.\ Lett.\ {\bf 90}, 116801 (2003).
\bibitem{Jac02} Ph. Jacquod, H. Schomerus, and C.W.J. Beenakker,
Phys.\ Rev.\ Lett.\ {\bf 90}, 207004 (2003).

\bibitem{Aga00} O. Agam, I. Aleiner, and A. Larkin, Phys.\ Rev.\ Lett.\ {\bf 85}, 3153 (2000).

\bibitem{Two03} J. Tworzyd\l o, A. Tajic, H. Schomerus, and C. W. J. Beenakker,
Phys.\ Rev.\ B {\bf 68}, 115313 (2003).

\bibitem{Ale96} I. L. Aleiner and A. I. Larkin, Phys. Rev. B {\bf 54}, 14423 (1996).
\bibitem{Ada03} I. Adagideli, Phys. Rev. B {\bf 68}, 233308 (2003). 

\bibitem{Gor03}  M.C. Goorden, Ph. Jacquod, and C.W.J. Beenakker, Phys. Rev. B {\bf 68}, 
220501 (2003).  

\bibitem{Bog92} E. B. Bogomolny, Nonlinearity\ {\bf 5}, 805 (1992).
\bibitem{Pra03} R. E. Prange,  Phys.\ Rev.\ Lett.\ {\bf 90}, 070401 (2003).

\bibitem{Bor91} F. Borgonovi, I. Guarneri, and D. L. Shepelyansky, Phys.\ Rev.\ A {\bf 43}, 4517 (1991).
\bibitem{Bor92} F. Borgonovi and I. Guarneri, J.\ Phys.\ A {\bf 25}, 3239 (1992).
\bibitem{Fyo00} Y. V. Fyodorov and H.-J. Sommers, JETP Lett.\ {\bf 72}, 422 (2000).
\bibitem{Oss02} A. Ossipov, T. Kottos, and T. Geisel, Europhys.\ Lett.\ {\bf 62}, 719 (2003).

\bibitem{Kor03} A. Kormanyos, Z. Kaufmann, C. J. Lambert, and J. Cserti, Phys. Rev. B {\bf 67}, 172506 (2003).

\bibitem{Jac03} Ph. Jacquod and E. V. Sukhorukov, cond-mat/0311528.

\end{thebibliography}
\end{document}